# Polarization entanglement by two simultaneous backward phase-matching processes in a single crystal

Ming-Yuan Gao[1,2,5], Yin-Hai Li[1,2,5], Zhao-Qi-Zhi Han[1,2], Qiang Zhou[1,4], Guang-Can Guo[1,2,3] Zhi-Yuan Zhou[1,2,3,*] and Bao-Sen Shi[1,2,3,*]

[1] CAS Key Laboratory of Quantum Information, University of Science and Technology of China, Hefei, Anhui 230026, China
[2] CAS Center for Excellence in Quantum Information and Quantum Physics, University of Science and Technology of China, Hefei 230026, China
[3] Hefei National Laboratory, University of Science and Technology of China, Hefei 230088, China
[4] Institute of Fundamental and Frontier Sciences, University of Electronic Science and Technology of China, Chengdu 610054, China
[5] These two authors contributed equally to this article.
*corresponding authors: zyzhouphy@ustc.edu.cn; drshi@ustc.edu.cn

Entanglement enables many promising applications in quantum technology. Devising new generation methods and harnessing entanglement are prerequisites for practical applications. Here we realize a distinct polarization-entangled source by simultaneously achieving type-0 and type-I backward quasi-phase matching (BQPM) through spontaneous parametric down-conversion in a single bulk crystal, which is different from all previous entangled-source configurations. Pumping the crystal with a single polarized beam generates a non-maximally polarization-entangled state, which can be further projected to a maximal Bell state with a pair of Brewster windows. Hong–Ou–Mandel interference experiments are done on polarization-degenerate photon pairs for both type-0 and type-I BQPM processes for the first time. The emitted photons in both processes have a bandwidth as narrow as 15.7 GHz. The high quality of this source is characterized by various methods. The rather simple configuration, narrow bandwidth, and high entanglement quality make the source very promising for many quantum information tasks.

Entanglement is indispensable for leading-edge applications in quantum information processing. Among various technical approaches to generate photonic entanglement states, spontaneous parametric down-conversion (SPDC) in nonlinear materials is one of the most effective and convenient methods in terms of fiber coupling efficiency, photon flux rate, and entanglement fidelity [1-5]. In SPDC, a strong pump laser beam with high frequency probabilistically splits into two daughter photons, commonly called the signal and idler. Though entanglement can be built up in various degrees of freedom, polarization entanglement is the most studied, and various generation configurations have been invented to achieve superior performances [6-15]. Two widely used configurations are the crossed-crystals scheme [7] and the Sagnac interferometer scheme [8, 9, 12-15]. In the crossed-crystals scheme, two crystals with the optical axis rotated by 90º are placed in sequence and pumped with a diagonally polarized laser, and the polarization entanglement is generated by two down-converters. In the Sagnac interferometer scheme, a single crystal is bidirectionally pumped in a polarization Sagnac interferometer. The popularity of these two schemes is due to the intrinsic stability and the fact that no active stabilization of the interferometer is required for the common path between three interacting light beams.

We have seen significant progress in improving SPDC sources during the past two decades, mostly concerning such parameters as tuning ability, efficiency, and purity. Especially, the advent and development of quasi-phase matching (QPM) offers great opportunities for engineering high-quality sources in new configurations. Highly efficient sources can be obtained with long QPM crystals [4, 5, 9-15] or waveguides [16-19]. In QPM, both forward and backward configurations can be realized. Most high-quality sources are based on forward QPM schemes [4, 5, 9-19] in which the three interacting waves propagate in the same direction. In the backward QPM (BQPM) configuration, the idler counterpropagates with respect to the pump and signal [20]. This configuration has been widely studied for second-harmonic generation [21] and mirrorless optical parametric oscillation [22]. Though photon pair generation in BQPM has been theoretically studied for decades [23-25], experimental progress has been achieved only in recent years [26-28]. Moreover, entanglements have not been generated and characterized in most experiments, except for a polarization entanglement achieved with type-II BQPM using a Sagnac interferometer [29]. With the distinct feature of BQPM, is there any new mechanism that can be used to generate polarization entanglement with the advantages of simplicity and robustness?

The answer is yes. Here we report a totally different polarization entanglement generation based on BQPM. In our scheme, both type-0 and type-I BQPM can be realized simultaneously in a periodically poled potassium titanyl phosphate (PPKTP) crystal, and a single linearly polarized pump laser can be used to generate a high-quality non-maximally polarization-entangled state with coefficients proportional to the square of the effective nonlinear coefficients. The state can be further projected to a maximally entangled Bell state using a pair of Brewster windows (BWs). Typical methods are used to characterize the quality of the entangled sources. We perform Hong–Ou–Mandel (HOM) interference experiments. The visibilities corresponding to type-0 and type-I without a spectral filter are $(84.9 \pm 1.1)$ % and $(84.6 \pm 4.2)$ % respectively, and they have the same bandwidth of 15.7 GHz. For the non-maximally entangled state, polarization interference at four bases indicates that all visibilities are greater than 98%, and quantum state tomography shows a fidelity of $0.980\pm0.002$. For the projected maximal Bell state, the average visibility at the four bases is greater than 96%, and the state fidelity is $0.947 \pm 0.009$. We also measure the Clauser–Horne–Shimony–Holt (CHSH) Bell inequality, which gives an S parameter of $S=2.725 \pm 0.107$. These results indicate this source is very promising for application in various quantum information processing tasks.

*Principles for the polarization-entangled source.* In our source configuration, the wavevector mismatch of QPM using PPKTP for backward-wave SPDC is given by [30]

$$\Delta k_0 = k_{pz} - k_{sz} + k_{iz} - k_m$$
$$\Delta k_1 = k_{pz} - k_{sy} + k_{iy} - k_m, \quad (1)$$

where $k_q = 2n_q\pi / \lambda_q$; $q = p$, $s$, and $i$, which are the wavevectors of the pump, forward propagating signal, and backward propagating idler respectively; $y$ and $z$ are the crystal axes corresponding to horizontal (H) and vertical (V) polarization; $k_m$ is the grating vector given by $k_m = 2\pi m/\Lambda$, where $m$ is the QPM order; $\Lambda$ is the poling period of the crystal; and $\Delta k_0$ and $\Delta k_1$ are the type-0 and type-I wavevector mismatches respectively. If the emitted photon pair is wavelength degenerate, the signal and idler photon wavevectors are canceled, and then the phase mismatching only depends on the wavevector $k_{pz}$. Therefore, both type-0 and type-I BQPM processes can be simultaneously phase-matched with the same $k_m$. The probability of a pump photon splitting into a signal and an idler photon corresponding to type-0 and type-I in SPDC is $P \propto d_{il}^2$ [31,32], where $d_{il}$ is the effective nonlinear coefficient corresponding to different QPM types. The effective nonlinear coefficients for type-0 and type-I BQPM are $d_0 = 2d_{33}/m\pi$ and $d_1 = 2d_{32}/m\pi$ respectively [32]. The entangled state generated in this process can be expressed as

$$|\Psi\rangle_{NM} = \frac{1}{\sqrt{R+1}}\left(|HH\rangle + e^{i\theta}\sqrt{R}|VV\rangle\right), \quad (2)$$

where $R = d_{33}^2/d_{32}^2$ is the ratio between the two phase-matching processes, which we have measured to be $R = 14.83$ [30]; and $\theta = (k_{sz} - k_{sy})L_c$ is the relative phase associated with the crystal birefringence, where $L_c$ is the crystal length. Therefore, $\theta$ can be compensated to 0 or $\pi$ using a KTP crystal with the same length but with its optical axes rotated by 90° with respect to the SPDC crystal. The non-maximally polarization-entangled state can be further manipulated to obtain a maximally entangled Bell state if both photon pairs generated from the type-0 phase matching are attenuated by $\sqrt{R}$, and photon pairs from the type-I phase matching are not attenuated at all. This attenuation functionality can be realized with a coated BW. If a pair of BWs is used, the non-maximally polarization-entangled state is transformed to

$$|\Phi\rangle_M = \frac{1}{\sqrt{2}}\left(|HH\rangle + e^{i\varphi}|VV\rangle\right), \quad (3)$$

where $\varphi$ is the relative phase, which includes the additional phase induced by the BWs. This phase can be finely tuned by controlling the temperature of the compensated KTP crystal. The working principle for our source is introduced above, and the performances of the source are detailed below.

*Experimental results.* The experimental setup is shown in Fig. 1. A single-mode fiber-(SMF-) coupled continuous-wave 778.33 nm laser beam was used to pump a bulk PPKTP crystal (Raicol Crystals, x-cut), which had dimensions of 1 mm× 2 mm× 4.5 mm, and 7th-order BQPM was achieved with a poling period of 2.95 μm. The degenerate type-0 and type-I phase-matching temperature of the crystal was 19.00 °C. The pump laser was focused on the center of the crystal with a 300 mm lens, and the beam waist at the center of the crystal was approximately 70 μm. The signal and idler photons generated were coupled to single-mode fibers (SMF2 and SMF3) with lenses. The pump laser beam was removed by long-pass filters before the photons were coupled into the single-mode fibers. The coupled photons were detected with superconductor nanowire single-photon detectors (SNSPDs), which had a detection efficiency of approximately 80% at 1550 nm. Finally, the coincidence between the photon pairs was measured with a time window of 1.6 ns (PicoQuant, TimeHarp 260).

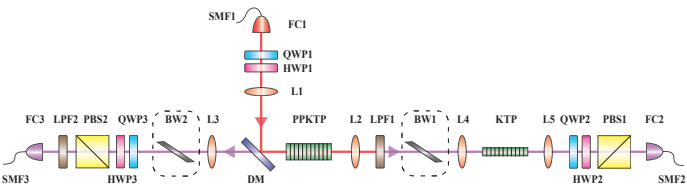

Fig. 1. Experimental setup. SMF: single-mode fiber; FC: fiber coupler; Q(H)WP: quarter-(half-) wave plate; PBS: polarization beam splitter; L: lens; DM: dichroic mirror; LPF: long-pass filter; BW: Brewster window; PPKTP: periodically poled KTP crystal.

To show the photon pairs generated in both types of BQPM were the same and obtain the bandwidth of the photon pair, HOM interference experiments were performed for photon pairs generated from both types of phase matching [33]. The detailed setup for HOM interference is shown in Fig. A1 in Appendix A. SMF2 was directly connected to one input of the fiber beam splitter (FBS), but there was a fiber-optic delay line between SMF3 and the other input of the FBS. The outputs of the FBS were connected with polarization controllers and finally connected to SNSPDs. FC2 was placed on a translation stage with a precision of 10 μm. The pump power was set at 100 mW, and its polarization was V. The HOM interference curves corresponding to type-0 and type-I were obtained by varying delays of the fiber-optic delay line and moving the translation stage, as shown in Fig. 2(a) and (b). Coincidence counts of the HOM dip were obtained by moving the translation stage only. The visibilities corresponding to type-0 and type-I were (84.9 ± 1.1) % and (84.6 ± 4.2) % respectively. They had an almost equal photon pair bandwidth of 15.7 GHz from the full width at half maximum (FWHM) of the fitted curves.

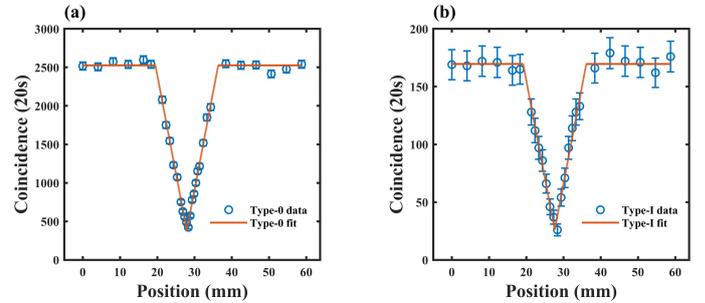

Fig. 2. Measured HOM interference curves corresponding to (a) type-0 and (b) type-I BQPM.

To characterize the non-maximally entangled state (the two BWs are removed in Fig. 1), we first measured polarization interference. A KTP crystal (Raicol Crystals, x-cut) was chosen that had the same length as the PPKTP crystal but with orthogonal optical axes. The KTP crystal was used to compensate for the relative phase, and its temperature was controlled. The pump power was set at 300 mW. HWP2 was set to 0°, 45°, −22.5°, and 22.5° to project one of the photon states in $|H\rangle$, $|V\rangle$, $|A\rangle$, and $|D\rangle$ respectively. Polarization interference curves were obtained by rotating HWP3, as shown in Fig. 3(a). The visibilities calculated from the fitted curves are (98.6 ± 0.4) %, (99.0 ± 0.1) %, (98.3 ± 0.2) %, and (98.4 ± 0.2) %.

Then quantum state tomography [34] was performed to obtain the density matrix $\rho_1$ of the generated state. The resulting reconstruction is shown in Fig. 3(b) and (c). Its fidelity relative to the ideal state is $0.980 \pm 0.002$. The ideal state here is given by

$$|\Psi_1\rangle = \frac{1}{\sqrt{R+1}}\left(|HH\rangle - \sqrt{R}|VV\rangle\right), \quad R = 14.83. \quad (4)$$

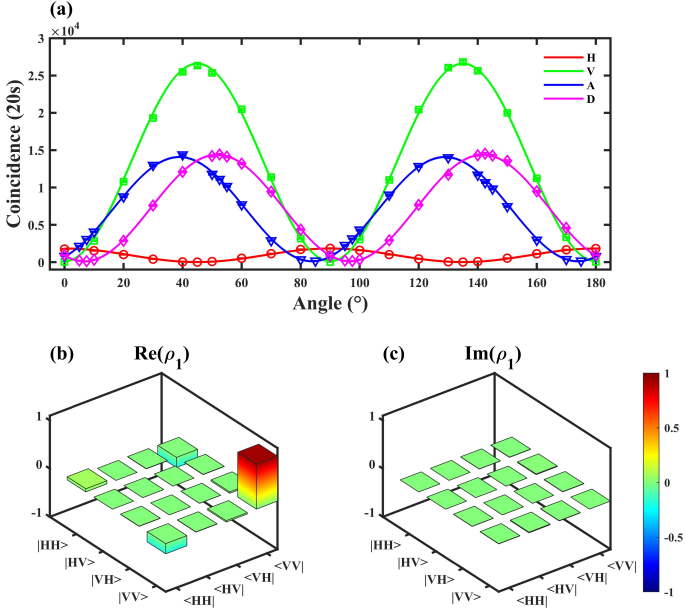

Fig. 3. Polarization entanglement characterization for the non-maximally entangled state. (a) Polarization interference in four bases; (b) and (c) are the reconstructed real and imaginary parts of the density matrix.

Then we used two BWs to transform the non-maximally entangled state to the maximally entangled state. The BWs in the experiment were positioned at Brewster's angle (approximately 56°). The ratio of the transmittance for H polarization to that for V polarization is about $\sqrt{R}$. To characterize the maximally entangled state, polarization interference was first measured. The visibilities were $(99.1 \pm 0.4)$ %, $(99.4 \pm 0.3)$ %, $(92.0 \pm 1.3)$ %, and $(93.6 \pm 1.2)$ % for the four bases $|H\rangle$, $|V\rangle$, $|A\rangle$, and $|D\rangle$ respectively, as shown in Fig. 4(a). To further obtain the density matrix $\rho_2$ of the maximally entangled state and the fidelity relative to the ideal Bell state, we performed quantum state tomography. The real and imaginary parts of the density matrix are shown in Fig. 4(b) and (c), and the fidelity relative to the Bell state $|\Phi^+\rangle = (|HH\rangle + |VV\rangle)/\sqrt{2}$ is $0.947 \pm 0.009$. We also performed the CHSH–Bell inequality test [35] and obtained S=$2.725 \pm 0.107$. Note that errors in our measurements are all estimated assuming Poisson statistics.

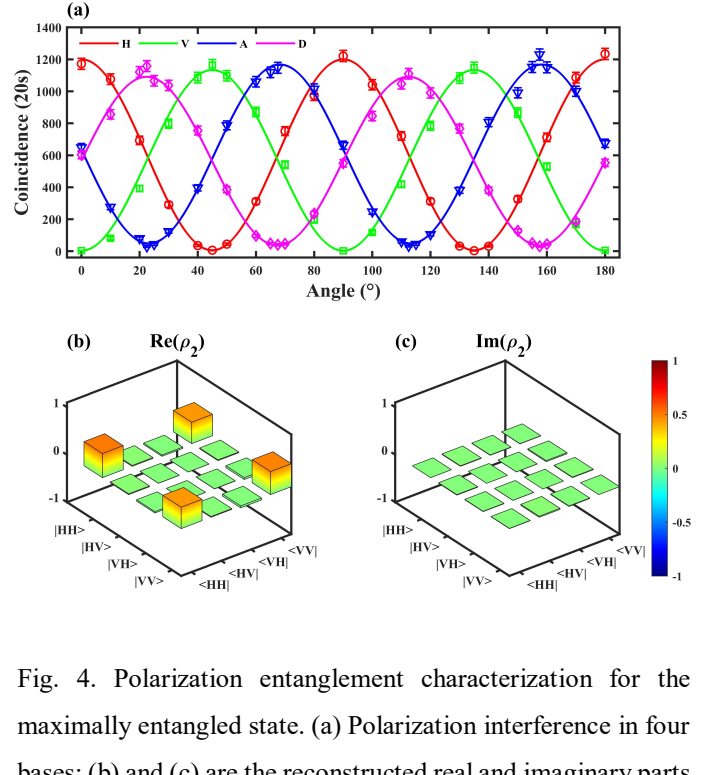

Fig. 4. Polarization entanglement characterization for the maximally entangled state. (a) Polarization interference in four bases; (b) and (c) are the reconstructed real and imaginary parts of the density matrix.

Finally, we give the estimated photon pair rate output from the crystal. The total detection efficiency of the signal and idler after considering the transmittance losses, coupling efficiency, and detection efficiency are 0.22 and 0.12 respectively, so the spectral brightness is approximately 60 $(s \cdot mW \cdot nm)^{-1}$.

*Discussion and Conclusions.* In Table 1, we compare the performance of our source with that of other works based on a bulk PPKTP or PPKTP waveguide. The result shows that our source has the same normalized brightness as the forward Sagnac-based sources and a narrower bandwidth. The single-pass configuration of our source guarantees it is more compact and stable. In addition, it may have unique application prospects in some situations. To illustrate the point, Sagnac-based sources rely on two-color polarization devices, which are not well developed and are expensive to process in certain special-purpose bands such as mid-infrared bands (for example, 3μm), and high-quality two-color polarization beam splitters are difficult to obtain. In this case, polarization-entangled sources in this band can be generated using our scheme without two-color polarization devices.

In conclusion, we generated narrowband photon pairs and entangled states at the telecom band by backward-wave SPDC in a bulk PPKTP crystal. Taking advantage of the simultaneous existence of type-0 and type-I BQPM in this crystal, we characterized the non-maximally entangled state with a certain coefficient directly generated by it. We projected the non-maximally entangled state to the maximally entangled state with a fidelity of $0.947 \pm 0.009$ by adding BWs. The source can generate collinear, degenerate, and separate signal and idler photon pairs, which cannot be realized in a forward QPM configuration. Our implementation provides a new perspective on generating polarization-entangled sources. The only defect is that high-order QPM processes are used, which greatly reduces the spectral brightness, but this can be overcome if the difficulty in fabricating a short poling period is solved in the future. Once first-order BQPM is used, the brightness of the source will be sufficiently increased. It is worth mentioning that the period poling technique developed by some researchers is sufficient to achieve first-order backward quasi-phase-matched SPDC photons [36]. The current source brightness can be further improved in the waveguide system and can be extended to PPLN and resonator sources. The present polarization-entangled source has great potential for use in various quantum information processing tasks.

Table 1. Comparison of different photon sources.

| Reference | Method | Wavelength (nm) | Bandwidth (GHz) | Brightness $[(s \cdot mW \cdot nm)^{-1}]$ | Normalized Brightness[a] |
|---|---|---|---|---|---|
| [12][b] | Forward bulk Sagnac interferometer | 810 | 137 | $3.4 \times 10^6$ | $2.51 \times 10^5$ |
| [13] | Forward bulk Sagnac interferometer | 1550 | 299 | $3.0 \times 10^4$ | $1.92 \times 10^4$ |
| [29] | Backward waveguide Sagnac interferometer | 1553.5 | 7.1 | $4.2 \times 10^5$ | -[c] |
| This work | Backward single-pass bulk configuration | 1556.6 | 15.7 | $6.0 \times 10^1$ | $1.91 \times 10^4$ |

[a] Details of the normalized brightness are provided in Appendix C.
[b] The spectral brightness here is obtained from 82000 photon pairs/s detected at 1 mW of pump power after considering the total loss.
[c] The effective mode area is not given in the text and therefore cannot be calculated.

This work is supported by the National Key Research and Development Program of China (2022YFB3607700, 2022YFB3903102), National Natural Science Foundation of China (NSFC) (11934013, 92065101, 62005068), Innovation Program for Quantum Science and Technology (2021ZD0301100), and the Space Debris Research Project of China (No. KJSP2020020202).

*Appendix A: Detail experimental setup for HOM interference measurements*

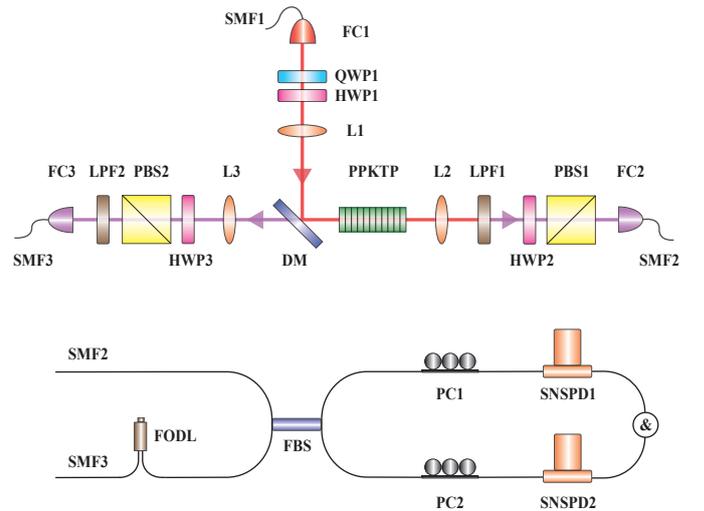

Fig. A1. Experimental setup of HOM interference experiments. SMF: single mode fiber; FC: fiber coupler; Q(H)WP: quarter (half) wave plate; PBS: polarization beam splitter; L: lens; DM: dichroic mirror; LPF: long-pass filter; PPKTP: periodically poled potassium titanyl phosphate crystal; FODL: fiber-optic delay line; FBS: fiber beam splitter; PC: polarization controller; SNSPD: superconductor nanowire single-photon detector.

*Appendix B: Additional information for measurements of non-maximally entangled states*

We perform polarization interference measurements on a pure state with the general form $|\Psi_0\rangle = c_1|H\rangle|H\rangle + c_2|V\rangle|V\rangle$, where $|c_1|^2 + |c_2|^2 = 1$. The setup of polarization interference measurements is shown in Fig. B1. The measurement of the photon state $|D\rangle$ is derived. HWP2 is set to $+22.5°$, polarization interference curves are obtained by rotation HWP1.

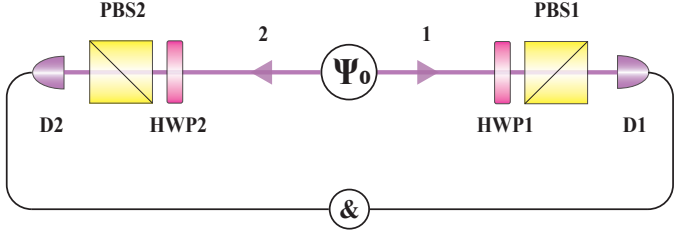

Fig. B1. Experimental setup of polarization interference measurements. HWP: half-wave plate; PBS: polarization beam splitter; D: detector.

The state after HWPs is given by

$$|\Psi_1\rangle = c_1 \frac{|H_1\rangle + |V_1\rangle}{\sqrt{2}} (\cos\theta|H_2\rangle + \sin\theta|V_2\rangle) \\ + c_2 \frac{|H_1\rangle - |V_1\rangle}{\sqrt{2}} (\sin\theta|H_2\rangle - \cos\theta|V_2\rangle), \quad (B1)$$

where $|H_i\rangle(|V_i\rangle)$, $i=1,2$, the subscripts refer to the two ports shown in Fig. B1, and the state is obtained by rotating HWP1 $\theta/2$ relative to $0°$. The state after polarization beam splitters is given by

$$|\Psi_2\rangle = \frac{1}{\sqrt{2}} c_1 \cos\theta |H_1\rangle|H_2\rangle + \frac{1}{\sqrt{2}} c_2 \sin\theta |H_1\rangle|H_2\rangle. \quad (B2)$$

The measuring result is given by

$$\left| \frac{1}{\sqrt{2}} c_1 \cos\theta + \frac{1}{\sqrt{2}} c_2 \sin\theta \right|^2. \quad (B3)$$

In the same way, the result of $|H\rangle$, $|V\rangle$, and $|A\rangle$ can be obtained.

Then we consider the state

$$|\Psi_0\rangle = \frac{1}{\sqrt{R+1}}(|HH\rangle + \sqrt{R}|VV\rangle), \quad R = 14.83. \quad (B4)$$

Using the derived results, the polarization interference curve is shown in Fig. B2, which is in good agreement with our experimental data.

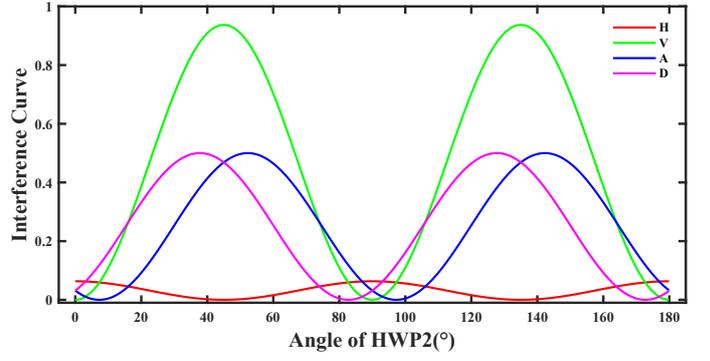

Fig. B2. Polarization interference curves of the given state.

*Appendix C: The normalized brightness*

Here we define normalized brightness to facilitate comparison of the brightness of the source under different parameters (beam waist, crystal length, and phase matching order). The normalized brightness is the spectral brightness obtained by further theoretical calculation using the brightness data assuming that the beam waist is 50 μm, the crystal length is 10 mm, and the matching order is the 1st. The unit of normalized brightness is (s·mW·nm)$^{-1}$. The spectral brightness has expressions B~1/A [31], ~$L^{3/2}$ [12], ~$1/m^2$ [32], where A is the effective area where the pump power acts (This can be understood as the power density of 1 mW), L is the length of the crystal, and m is the matching order. Therefore, the normalized brightness $B_n$ can be calculated by the following formula

$$B_n = \left(\frac{A}{A_0}\right) \cdot \left(\frac{L_0}{L}\right)^{3/2} \cdot \left(\frac{m}{m_0}\right)^2 \cdot B,$$

where $A_0 = \pi \cdot (50\mu m)^2$, $L_0 = 10\text{mm}$, $m_0 = 1$.